# Room-temperature magnetoelectric effect in lead-free multiferroic (1-x) $Ba_{0.95}Ca_{0.05}Ti_{0.89}Sn_{0.11}O_3$ - (x) $CoFe_2O_4$ particulate composites

Youness Hadouch[1,2,*], Daoud Mezzane[1,2], M'barek Amjoud[1], Nouredine Oueldna[3], Yaovi Gagou[2], Zdravko Kutnjak[4], Valentin Laguta[5,6], Yakov Kopelevich[7], Khalid Hoummada[3], Mimoun El Marssi[2].

*1 Laboratory of Innovative Materials, Energy and Sustainable Development (IMED), Cadi- Ayyad University, Faculty of Sciences and Technology, BP 549, Marrakech, Morocco.*

*2 Laboratory of Physics of Condensed Matter (LPMC), University of Picardie Jules Verne, Scientific Pole, 33 rue Saint-Leu, 80039 Amiens Cedex 1, France.*

*3 Aix-Marseille University - CNRS, IM2NP Faculté des Sciences de Saint-Jérôme case 142, 13397 Marseille, France*

*4 Jozef Stefan Institute, Jamova Cesta 39, 1000 Ljubljana, Slovenia.*

*5 Institute of Physics AS CR, Cukrovarnicka 10, 162 53 Prague, Czech Republic.*

*6 Institute for Problems of Materials Science, National Ac. of Science, Krjijanovskogo 3, Kyiv 03142, Ukraine.*

*7 Universidade Estadual de Campinas-UNICAMP, Instituto de Física "GlebWataghin", R. Sergio Buarque de Holanda 777, 13083-859 Campinas, Brazil.*

*Corresponding author:

E-mail: hadouch.younes@gmail.com; youness.hadouch@etud.u-picardie.fr

ORCID: https://orcid.org/0000-0002-8087-9494

## *Highlights:*

- Composite multiferroic (1-x) $Ba_{0.95}Ca_{0.05}Ti_{0.89}Sn_{0.11}O_3$ – (x) $CoFe_2O_4$ ceramics were prepared by mechanical mixing method;
- Ferroelectric and ferromagnetic hysteresis loops indicate the multiferroic behavior of composites at room temperature;
- 0.7 BCTSn-0.3 CFO composition shows a maximum magnetoelectric.

## Abstract:

Multiferroic particulate composites (1-x) $Ba_{0.95}Ca_{0.05}Ti_{0.89}Sn_{0.11}O_3$–(x) $CoFe_2O_4$ with (x = 0.1, 0.2, 0.3, 0.4 and 0.5) have been prepared by mechanical mixing of the calcined and milled individual ferroic phases. X-ray diffraction and Raman spectroscopy analysis confirmed the formation of both perovskite $Ba_{0.95}Ca_{0.05}Ti_{0.89}Sn_{0.11}O_3$ (BCTSn) and spinel $CoFe_2O_4$ (CFO) phases without the presence of additional phases. The morphological properties of the composites were provided by using Field Emission Scanning Electron Microscopy. The BCTSn-CFO composites exhibit multiferroic behavior at room temperature, as evidenced by ferroelectric and ferromagnetic hysteresis loops. The magnetoelectric (ME) coupling was measured under a magnetic field up to 10 kOe and the maximum ME response found to be 0.1 mV cm$^{-1}$ Oe$^{-1}$ for the composition 0.7 BCTSn-0.3 CFO exhibiting a high degree of pseudo-cubicity and large density.

*Acknowledgements:*

This research is financially supported by the European Union Horizon 2020 Research and Innovation actions MSCA-RISE-ENGIMA (No. 778072), MSCA-RISE-MELON (No. 872631), FAPESP and CNPq, Brazilian agencies.

*Formatting of funding sources:*

The European Union's Horizon 2020 research;
MSCA-RISE-ENGIMA (No. 778072)
MSCA-RISE-MELON (No. 872631).

*1 Introduction:*

The microelectronics industry has developed progressively compact integrated circuits with even more sophisticated functionalities in response to the rising demand for electronic devices, whether portable or inserted into other systems. This complexity encourages the development of multifunctional materials, which combine well-known features into a single component [1]–[3].

Researchers have become interested in multiferroic materials exhibiting simultaneous ferroelectric and ferromagnetic orderings because of their attractive physical characteristics and potential uses in spintronics, data storage, and sensors. All of these applications are based on the magnetoelectric effect (ME), which refers to the coupling of a material's electric and magnetic properties [4]–[6]. A gendered magnetic field results from an applied electric field that distorts the ferroelectric/piezoelectric phase, which in turn distorts the magnetostrictive/magnetic phase. However, this coupling phenomenon can be reversed [7], [8], and the ME effect for composites may be stated as follows [9]:

$$ME_{effect} = \frac{magnetic}{mechanical} \times \frac{mechanical}{electrical} \ or \ \frac{electrical}{mechanical} \times \frac{mechanical}{magnetic} \quad Eq.\ (1)$$

The multiferroic materials are often classified into single-phase[10], [11] and composites[5]. Single-phase materials with intrinsic ME coupling are understudied because of their weak ME effect at or near room temperature, which restricts the applications [12]. Indeed, the ferroelectric mechanism, which requires formally empty d orbitals, in contrast to the ferromagnetic mechanism, which requires partially filled d orbitals[12], [13]. On the other hand, synthetic multiferroic composites with various forms of connectivity, such as particulate ceramic (0–3) [14], laminates/films (2-2) [15], [16], and rod/fibers core–shell (1-3) [17], exhibit an excellent extrinsic ME coupling created by an elastic ferroelectric-ferromagnetic interfacial interaction combining high magnetostriction and a large piezoelectric response. On the other hand, particulate-type composites are simple to make and have a uniform ME effect in all directions[14].

Lead (Pb) based materials are the most often employed ferroelectric phase in multiferroic composites because of their excellent polarization and piezoelectric properties [15]. However, lead-based materials have negative impact on the environment. Therefore, it is necessary to develop new environmentally friendly materials. $Ba(Ti_{0.8}Zr_{0.2})O_3$-$(Ba_{0.7}Ca_{0.3})TiO_3$ (BCZT) is the most studied piezoelectric material with a values of piezoelectric coefficient $d_{33}$ in the range of 500-600 pC $N^{-1}$ [16],[17]. Due to its high piezoelectric coefficient of $d_{33}$ =670 pC $N^{-1}$ and the morphotropic phase boundary (MPB) in $(Ba_{0.95}Ca_{0.05})(Ti_{0.89}Sn_{0.11})O_3$ (BCTSn), Zhu et al. demonstrated that this material might be used as an alternative for BCZT in piezoelectric applications[18]. Moreover, we previously reported that $Ba_{0.95}Ca_{0.05}Ti_{0.89}Sn_{0.11}O_3$ exhibited an enhanced energy storage performance of $W_{rec}$ = 124 mJ $cm^{-3}$ with a storage efficiency of $\eta$ = 91.07 % and a significant electrocaloric response of $\Delta T$ = 0.807 K [19]. On the other hand, due to its high magnetostriction coefficient ($\lambda \sim -110\times10^{-6}$), moderate saturation magnetization, and strong chemical stability, spinel ferrite $CoFe_2O_4$ (CFO) is the most commonly utilized magnetostrictive phase in magnetic materials [20].

In addition, the most studied type (0-3) multiferroic composites in the literature are based on combining of the piezoelectric phase BCZT and the magnetostrictive phase CFO. For example, Praveen et al studied multiferroic and magnetoelectric properties of [(1-x) $(Ba_{0.85}Ca_{0.15})(Zr_{0.1}Ti_{0.9})O_3$- (x) $CoFe_2O_4$] (weight fraction; x=0, 0.1, 0.2, 0.3, 0.4, 0.5 and 1) ceramic particulate composites [21] and Rani et al reported the magnetic and magnetocapacitance properties of (x) $CoFe_2O_4$ - (1-x)(0.5Ba$(Zr_{0.2}Ti_{0.8})O_3$-0.5$(Ba_{0.7}Ca_{0.3})TiO_3$)[22].

In this paper, we have prepared a newly-developed (0-3) magnetoelectric composites by combining the $Ba_{0.95}Ca_{0.05}Ti_{0.89}Sn_{0.11}O_3$ (BCTSn) and $CoFe_2O_4$ (CFO). The structural, morphological and multiferroic properties of the (1-x) $Ba_{0.95}Ca_{0.05}Ti_{0.89}Sn_{0.11}O_3$–(x)$CoFe_2O_4$

with (x = 0.1, 0.2, 0.3, 0.4 and 0.5) composites have been systematically investigated as a function of weight fraction (x).

## 2 Material and methods

### 2.1 Chemical synthesis:

Ferroelectric $(Ba_{0.95}Ca_{0.05})(Ti_{0.89}Sn_{0.11})O_3$ (BCTSn) and ferromagnetic $CoFe_2O_4$ (CFO) were synthesized using sol-gel and sol-gel self-combustion methods, respectively. The preparation of both pristine BCTSn and CFO has been thoroughly described in our previous work[19], [20]. Polycrystalline (1-x) BCTSn – (x) CFO particulate composites with different weight fractions x (x = 0.1, 0.2, 0.3, 0.4 and 0.5) were fabricated by mechanical mixing of the calcined and milled individual ferroic phases of BCTSn and CFO and then pressed into cylindrical pellets and sintered at 1300°C for 4 hours. Finally, the pellets were coated with silver paste on both surfaces for electrical and magnetoelectrical measurements.

### 2.2 Characterization

The XRD patterns of (1-x) BCTSn – (x) CFO ceramic composites were obtained by X-ray diffraction using the Panalytical X-Pert Pro with Cu-K$\alpha$ radiation ($\lambda$ = 1.54059 Å) at room temperature. The grain morphology of ceramics was analyzed using the scanning electron microscopy (SEM) in a HELIOUS 600 nanolab setup from FEI. The Raman spectra were recorded using a micro-Raman Renishaw spectrometer equipped with a CCD detector. Magnetic properties were measured using a Physical Property Measurement System (PPMS-DynaCool) Quantum Design apparatus that operated at room temperature under a magnetic field range of 0–2.5 kOe. The polarization-electric field (P-E) hysteresis loops were performed by CPE1701, PloyK, USA, with a high-voltage power supply (Trek 609-6, USA) at room temperature and 20 Hz. Magnetoelectric measurements were done using conventional Bruker EPR spectrometer EMX plus operating at 400-600 Hz frequency range at room temperature.

## 3. Results and discussion:

### 3.1 Structural and morphological analysis:

The Collected XRD data of pristine BCTSn, CFO and the composite systems[(1-*x*) BCTSn-(*x*) CFO with *x* = 0.1, 0.2, 0.3, 0.4 and 0.5] sintered at 1300°C for 4 hours are shown in Fig.1. It is apparent from the XRD patterns that all diffraction peaks of ferroelectric (black line) and ferrite (red line) phases appear in the composite samples without the presence of any impurities, proving that there was no chemical reaction between the ferroelectric and ferrite phases during the sintering process.

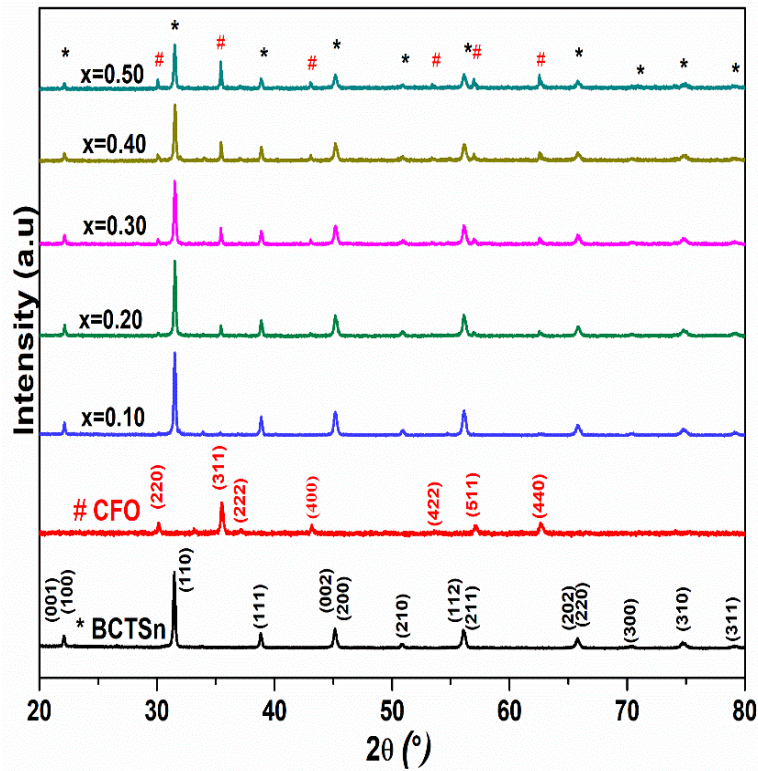

**Fig. 1** XRD patterns of the (1-x) BCTSn - (x) CFO composites.

Fig. 2 shows the structural refinement of the composite systems [(1-*x*) BCTSn- *x* CFO with *x* = 0.1, 0.2, 0.3, 0.4 and 0.5] using the software FULLPROF. The structural refinement of both pristine $Ba_{0.95}Ca_{0.05}Ti_{0.89}Sn_{0.11}O_3$ and $CoFe_2O$ was discussed in our previously reported works [19], [20].

The results revealed the coexistence of two distinct phases: BCTSn (*Amm2 & P4mm*) and CFO (*Fd$\bar{3}$m*), and no peak has been identified for other crystal symmetries. **Table 1** lists the corresponding ratios of all the phases, the results are in good agreement with the weight percentage ratios used in the synthesis of composites. This proves that the di-phase multiferroic composites were successfully synthesized. It should be noted that the lattice constants of the spinel and perovskite phases change from one composition to another. This could be due to strain created at the multiferroic phase boundary caused by the crystal structure mismatch[14]. Fig.2f shows the lattice distortion (c/a) of the orthorhombic and tetragonal phases in BCTSn. In general, the variation of lattice distortion (c/a) in ferroelectric could be correlated with the magnetoelectric coupling of the multiferroic composite [14]. The crystal is relatively close to pseudo-cubic symmetry if the lattice distortion (c/a) in both ferroelectric phases (orthorhombic and tetragonal) is reduced, which decreases the strain and improves domain switching [14]. In our case, the reduced value of (c/a) in both ferroelectric phases was noticed at x=0.3. Accordingly, the composite with the composition of 0.7 BCTSn and 0.3 CFO exhibited the highest degree of pseudocubicity.

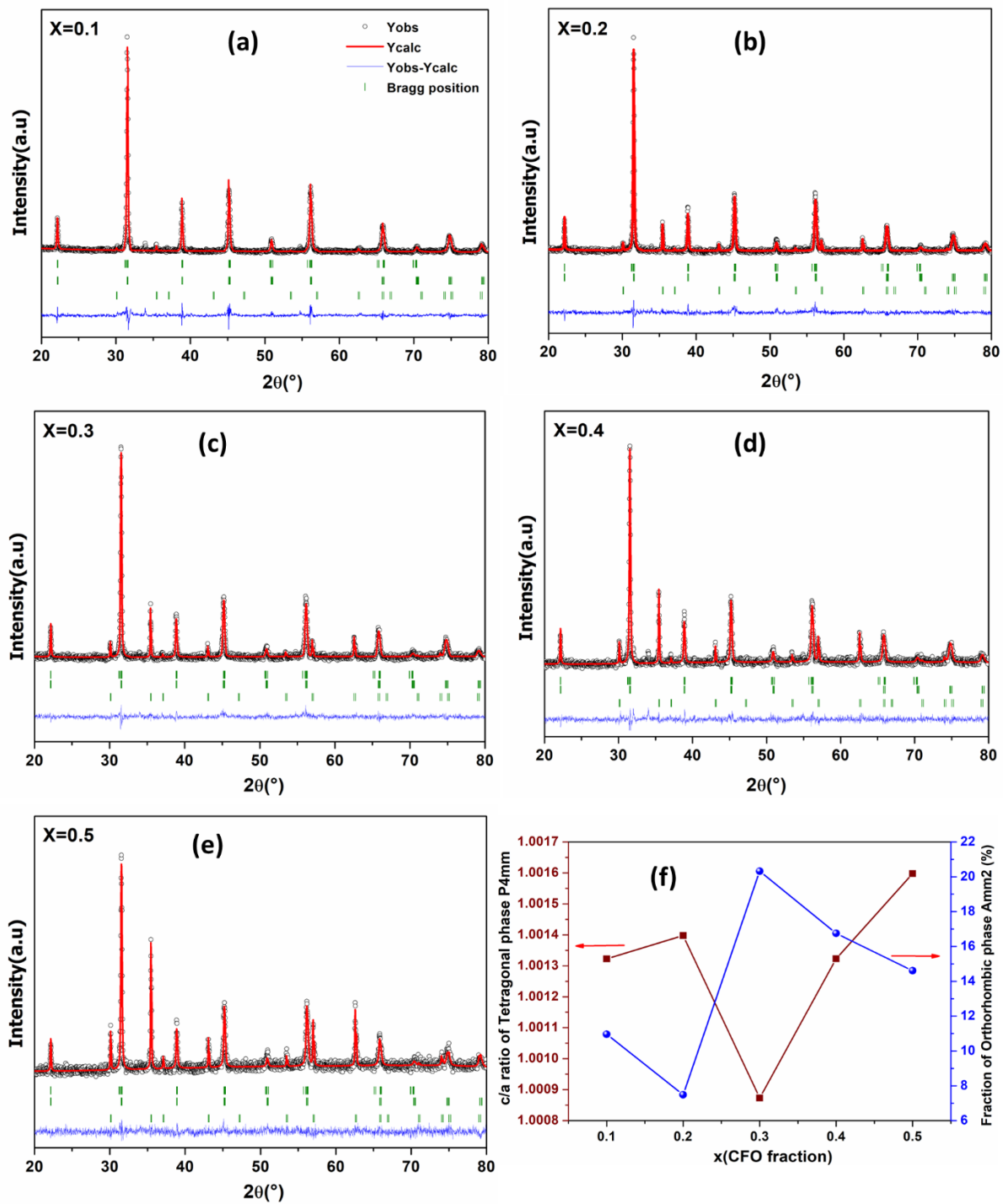

**Fig. 2 (a-e)** The Rietveld refined XRD patterns for the composites: (1-x) BCTSn - (x) CFO, **(f)** Variation of lattice distortion and % fraction of crystal phase Amm2 with CFO content in (1-x) BCTSn-(x) CFO composites.

Table 1 Rietveld refined XRD parameters for (1-x) BCTSn - (x) CFO composite samples at room temperature

| Composite (1-x) BCTSn - (x) CFO | Phase | Lattice parameters | | | Volume (Å)$^3$ | Fraction % |
|---|---|---|---|---|---|---|
| | | a (Å) | b (Å) | c (Å) | | |
| x=0.1 | $Fd\bar{3}m$ | 8.3721 | - | - | 586.81772 | 8.95 |
| | Amm2 | 4.0040 | 5.7261 | 5.7261 | 131.28404 | 10.96 |
| | P4mm | 4.0069 | 4.0069 | 4.0122 | 64.41686 | 80.08 |
| x=0.2 | $Fd\bar{3}m$ | 8.3792 | - | - | 588.31195 | 18.27 |
| | Amm2 | 4.0069 | 5.7273 | 5.7273 | 131.43419 | 7.49 |
| | P4mm | 4.0057 | 4.0057 | 4.0113 | 64.36385 | 74.23 |
| x=0.3 | $Fd\bar{3}m$ | 8.4002 | - | - | 592.74634 | 28.89 |
| | Amm2 | 4.0097 | 5.7273 | 5.7273 | 131.52276 | 20.32 |
| | P4mm | 4.0089 | 4.0089 | 4.0124 | 64.4844 | 50.79 |
| x=0.4 | $Fd\bar{3}m$ | 8.3854 | - | - | 589.61884 | 41.35 |
| | Amm2 | 4.0102 | 5.7278 | 5.7278 | 131.56541 | 16.75 |
| | P4mm | 4.0064 | 4.0064 | 4.0117 | 64.39276 | 41.90 |
| x=0.5 | $Fd\bar{3}m$ | 8.3949 | - | - | 591.62509 | 48.86 |
| | Amm2 | 4.0108 | 5.7276 | 5.7281 | 131.59231 | 14.61 |
| | P4mm | 4.0053 | 4.0053 | 4.0117 | 64.35741 | 36.53 |

To further investigate the (1-x) BCTSn-(x) CFO composite structure, Raman spectra were recorded in the frequency range of 100-900 cm$^{-1}$ (Fig. 3). The optical active Raman modes for CFO ($A_{1g} + E_g + 3T_{2g}$) are identified at position peaks (~212, ~314, ~474, ~581, ~619 and ~696 cm$^{-1}$), resulting from the motion of oxygen ions and both tetrahedral and octahedral sites ions in CFO spinel structure [23]. The modes at ~619 and ~696 cm$^{-1}$ correspond to the symmetric stretching vibration of FeO$_4$ tetrahedral sub lattice, and those at ~314 and ~581 cm$^{-1}$ depict the symmetric and asymmetric bending of the oxygen anions in the octahedral sublattice respectively. The second $T_{2g}(2)$ mode at ~474 cm$^{-1}$ corresponds to asymmetric stretching of the Fe-O (Co-O) bonds, while the $T_{2g}(3)$ at ~212 cm$^{-1}$ is ascribed to the translational shift of the whole FeO$_4$ tetrahedron [20]. For BCTSn phase, 4 A and 2 E modes are observed at 166, 195, 264, 303, 519, 722 cm$^{-1}$ [24]. The Peaks on the lower-frequency side (< 300 cm$^{-1}$) are associated with Ba/Ca-O vibrations, while those on the higher wavenumber side correspond to Ti/Sn-O vibrations [24]. The obtained Raman spectra for all compositions constitute Raman modes of the pure BCTSn and CFO phases individually without any strange Raman bands[25]. These results support the successful preparation of (0-3) type connectivity for BCTSn-CFO multiferroic composite materials.

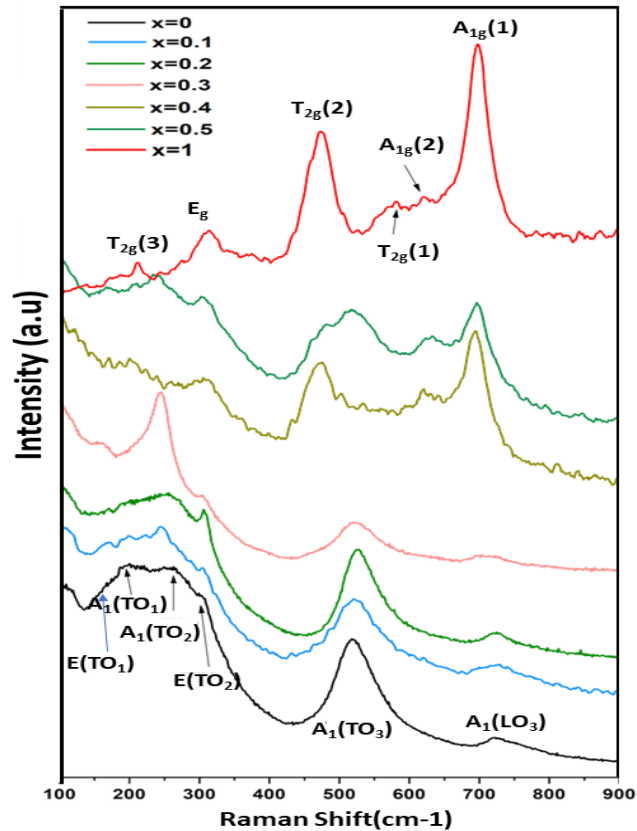

**Fig. 3** Raman spectra of CFO, BCTSn and (1-x) BCTSn - (x) CFO composites

Fig. 4 (a) shows the SEM micrograph of sintered pellets of 0.7 BCTSn – 0.3 CFO multiferroic composites. Due to the difference in molecular weights, BCTSn grains seem brighter while CFO grains appear darker gray. The density of the sintered composites increases with increasing CFO fraction reaching a maximum of 5.5 at x=0.3, and then remains constant (Fig. 4b). This can be attributed to the fact that CFO exhibits greater densification characteristics than BCTSn.

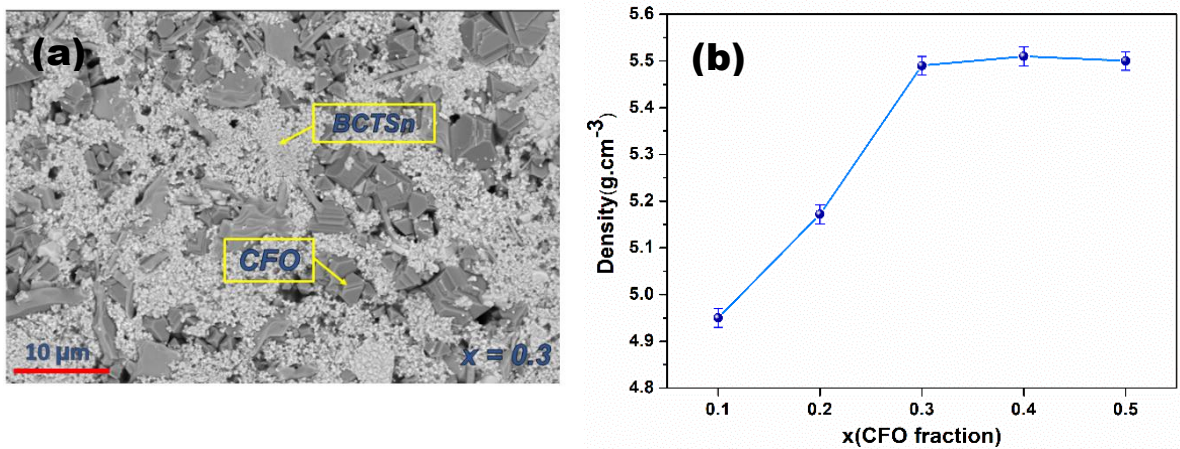

**Fig. 4 (a)** SEM micrographs of 0.7 BCTSn – 0.3 CFO, **(b)** bulk density as a function of ferrite content of various (1-x) BCTSn - (x) CFO composites.

## 3.2 Magnetic properties of magnetoelectric composites:

Fig.5 shows the room temperature magnetization vs magnetic field (M-H) hysteresis loops of the particulate composites (1-x) BCTSn - (x) CFO and that of pure CFO in the inset. All the composites' obtained hysteresis loops showed identical behavior to pristine CFO, confirming the materials' ferromagnetic composition. Different magnetic parameters, including magnetic coercive field ($H_c$), saturation magnetization ($M_s$) and remanent magnetization ($M_r$), are obtained and listed in table 2. Because of the decrease in the magnetic moment per unit volume caused by the presence of the non-magnetic BCTSn phase, $M_s$ and $M_r$ value both reduce systemically as the BCTSn content rises. $CoFe_2O_4$ coercivity is influenced by the form and structure of the grains, as was described in our earlier work [20]. Therefore, as CFO content rises, the increase in coercivity is most likely caused by the formation of $CoFe_2O_4$ clusters and interconnected structures. To compare the magnetic properties of the composites to the original magnetic phase of CFO, the reduced magnetization ($M_r/M_s$) of the individual composite was calculated (Table 2). It has been noticed that the values of reduced magnetization vary around one percent (1%), suggesting that there is no significant chemical diffusion or interaction between BCTSn and CFO phases since the chemical reactions can modify the original magnetic behavior of CFO [14].

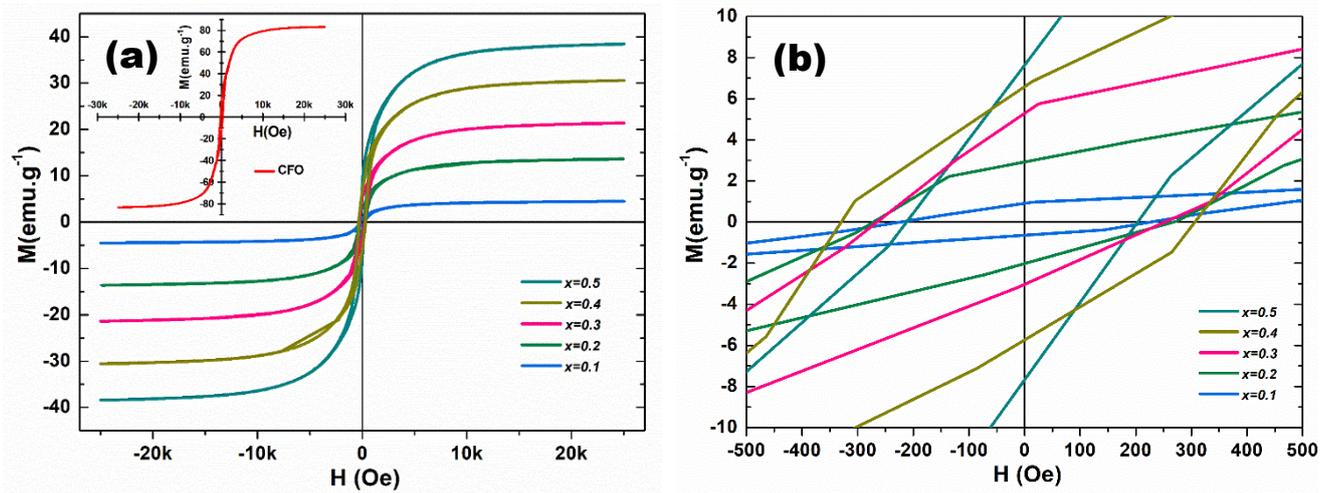

**Fig.5 (a)** Magnetic hysteresis loops of the particulate composites (1-x) BCTSn - (x) CFO with x= 0.10, 0.20, 0.30, 0.40, and 1.0 measured at room temperature (Inset shows M–H curve of pure CFO). **(b)** Enlarged view of room temperature M–H curve of all compositions.

Table 2. Different magnetic parameters are determined using the hysteresis loop $H_c$, $M_r$, $M_s$, and $M_r/M_s$ for different compositions.

| Composite (1-x) BCTSn - (x) CFO | $H_c$ : Magnetic coercive field (Oe) | $M_r$ : Remanent magnetization (emu g$^{-1}$) | $M_s$ : Saturation magnetization (emu g$^{-1}$) | $M_r/M_s$ Reduced magnetization |
|---|---|---|---|---|
| *Pure CFO*[20] | 284 | 14.96 | 83.00 | 0.1802 |
| *x=0.5* | 203.81 | 7.38 | 38.41 | 0.1921 |
| *x=0.4* | 305.66 | 6.18 | 30.61 | 0.2018 |
| *x=0.3* | 253.35 | 4.23 | 21.38 | 0.1978 |
| *x=0.2* | 268.58 | 2.72 | 13.66 | 0.1991 |
| *x=0.1* | 228.94 | 0.88 | 4.53 | 0.1942 |

### *3.3 Ferroelectric properties of magnetoelectric composites*

The polarization (P)-electric field (E) hysteresis loops of all BCTSn-CFO composites including, that of pure BCTSn under a maximum applied electric field of 20 kVcm$^{-1}$ are shown in Fig. 6. The ferroelectric hysteresis loops were measured at a frequency of 10 Hz. The obtained hysteresis loops confirm the ferroelectric nature of all composites. In addition, electrcic coercive field ($E_c$), maximal polarization ($P_{max}$) and remanent polarization ($P_r$) were extracted from Fig.6 and listed in table 3. It should be noted that compared to BCTSn, all composites exhibit lower Pr and higher Ec. This can be explained by the lossy nature of the composites, which increases when conductive ferrite particles are added. It is also worth noting that the values of $P_{max}$ decrease as CFO content in composites increases. In reality, the electric dipoles in ferroelectric systems are arranged in a specific order. However, due to the presence of ferrites, the long-range ordering of these electric dipoles is disrupted, thus reducing the electric polarization. Furthermore, the increased CFO concentration around BCTSn particles in the composite inhibits the movement of domain walls within BCTSn grains. As a result, when a strong electric field is applied, the BCTSn matrix cannot generate dipoles, reducing of ferroelectric polarization [14], [26].

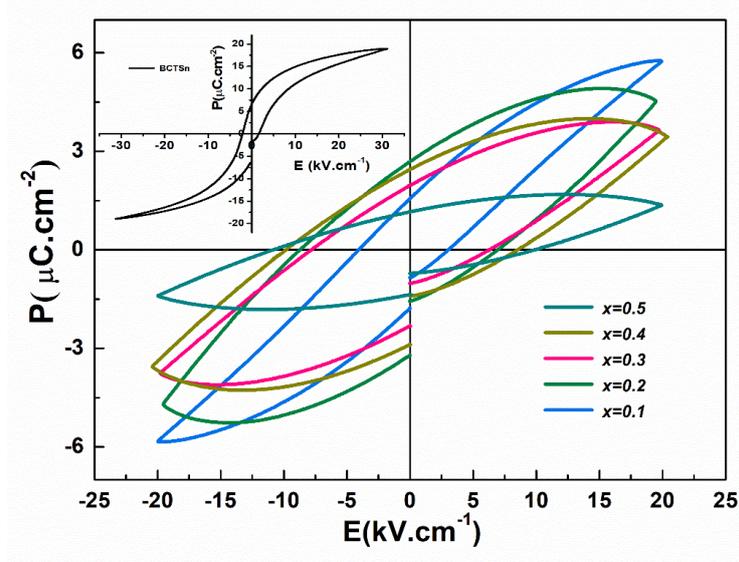

**Fig.6** Ferroelectric hysteresis loops of the particulate composites (1-x) BCTSn - (x) CFO with x= 0.10, 0.20, 0.30, 0.40, and 0.0 at room temperature (Inset shows P–E curve of pure BCTSn).

Table 3. Different magnetic parameters determined using the hysteresis loops. $E_c$, $P_r$ and $P_{max}$ for different compositions.

| Composite (1-x) BCTSn - (x) CFO | $E_c$ : Coercive field (kV.cm$^{-1}$) | $P_r$ : Remanent polarization (µC.cm$^{-2}$) | $P_{max}$ : Maximal polarization (µC.cm$^{-2}$) |
|---|---|---|---|
| x=0[19] | 1.68 | 6.77 | 18.93 |
| x=0.1 | 3.03 | 1.55 | 5.74 |
| x=0.2 | 6.95 | 2.67 | 4.91 |
| x=0.3 | 6.26 | 1.95 | 4.01 |
| x=0.4 | 8.53 | 2.44 | 3.97 |
| x=0.5 | 9.84 | 1.16 | 1.69 |

**Fig.7** shows how the composition x influences the magnetic and ferroelectric characteristics ($M_s$) and ($P_{max}$). Accordingly, our particle composites (1-x) BCTSn- x CFO have multiferroic characteristics.

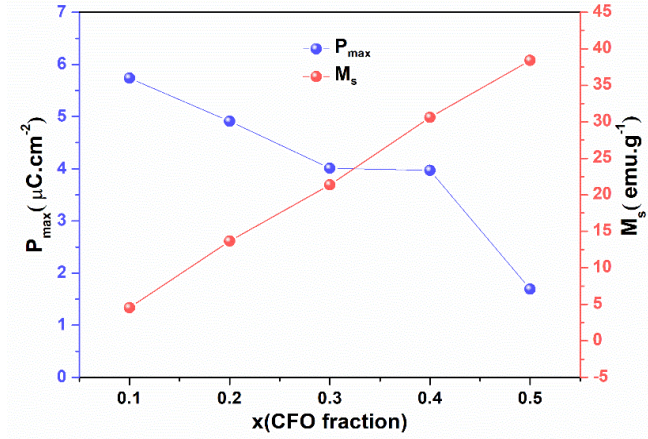

**Fig .7** Variation of $P_{max}$ and $M_s$ of the composites (1-x) BCTSn - (x) CFO with x = 0.10, 0.20, 0.30, 0.40 and 0.50 at 100 Hz frequency at the room temperature.

### *3.4 Magnetoelectric properties*

The ME effect can be described by the ME voltage coefficient($\alpha_{ME}$), which can be calculated by the following relation [27]:

$$\alpha_{ME} = \left(\frac{\partial E}{\partial H}\right) = \frac{V}{d \times H} \quad Eq.\ (2)$$

Where V is the induced ME voltage across the sample, d represents the thickness of the sample and H is the applied magnetic field.

Magnetoelectric coupling was studied for the (1-x) BCTSn- (x) CFO samples at room temperature using a dynamical method [28]. The sample was firstly poled at 295 K in the field of 20-25 kV cm$^{-1}$. Then the magnetoelectric current or voltage induced by a weak *ac* magnetic field of $H_{ac}$ = 1.8 Oe at frequency 400-600 Hz was measured at room temperature as a function of the *dc* bias field utilizing a high-sensitive lock-in-amplifier [28], [29]. Both magnetic fields were applied normal to the surface of the sample. In every experiment, more than two runs were repeated with a reversed direction of *dc* magnetic field, resulting in a change of signal sign. In this way, a possible spurious electromagnetic induction signal was distinguished from a true ME whose sign depends on the polarization×magnetic field product [6].

The measured data for four samples (x=0.1; 0.2; 0.3, and 0.4) are shown in Fig. 8. The ME response increases with the increase of the CFO magnetic content up to x=0.3, then it sharply decreases at further the growth of the CFO phase (inset fig.8). This is related to the fact that at x>0.4, the sample cannot be well poled and there excises some optimal ratio between ferroelectric and magnetic components [5]. Indeed, increasing the CFO content (above 30%) reduces the composite's overall resistivity and charge leakage due to an increase in migratory electrons along the interface between the two phases. Consequently, electric poling becomes more difficult to create dipole by applying a magnetic field, resulting in a reduction in ME coupling in composites [14]. The ME coefficient has a maximum at the field of about 2.5 kOe related to the maximal values of the derivative of the strain on *H* value and approximately coincides with the maximum of the d*M*/d*H* (i.e. magnetic susceptibility) value. There is also a weak hysteresis in the ME response attributed to hysteresis in the M-H loops.

This ME coupling is explained by the fact that the magnetic field creates a strain in the ferrite phase, and the strain causes stress in the ferroelectric phase due to the mechanical interactions of the ferrite and perovskite phases. Because stress promotes polarization, a voltage is produced in the grains [30]. As a result, the area of the interface between ferroelectric and ferromagnetic grains is critical in this type of strain-mediated ME coupling [31].

If we compare our data with similar data published for $BaTiO_3$ – $CoFe_2O_4$ ceramic composites, one can notice that the ME voltage coefficient in the studied by us BCTSn – CFO composites is weaker than that in the $BaTiO_3$ – $CoFe_2O_4$ system where it is mainly in the range 0.14 - 0.2 mV.cm$^{-1}$ Oe$^{-1}$ but strongly depends on method the of preparation [31]. Other comparable results on multiferroic composites have been published by various research groups (Table 4). It is expected that the efficiency of ME coupling in BCTSn – CFO ceramic composites can be substantially enhanced through variation of the sizes and shapes of the grains of FE and FM phases as well as the synthesis of core-shell nanostructures.

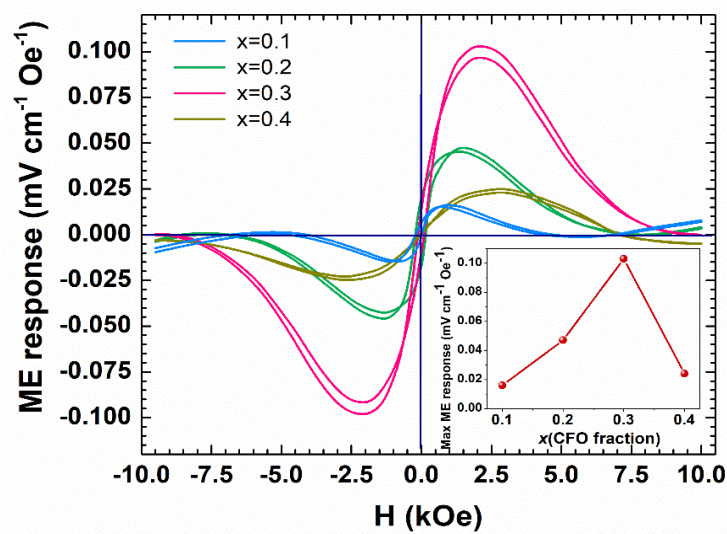

**Fig. 8** Amplitude of the ME coupling coefficient for the samples with x=0.1, 0.2, 0.3, and 0.4 as a function of the *dc* magnetic field.

Table 4. Comparison of Maximum ME response of various magnetoelectric composites.

| Material | Connectivity | Applied magnetic field (kOe) | Maximum ME response (mV cm$^{-1}$ Oe$^{-1}$) | References |
|---|---|---|---|---|
| 0.70BCTSn-0.30CFO | 0-3 | 2.1 | 0.10 | This work |
| 0.62BST-0.38CFO | 3-3 | 1.5 | 0.05 | [27] |
| 0.85BT-0.25CFO | 0-3 | -- | 0.14 | [32] |
| 0.80 (PMN-PT)- 0.20 (CFO) | 0-3 | 2.5 | 0.17 | [33] |
| 0.70 BT – 0.30 CFO | 0-3 | 0.5 | 0.12 | [34] |
| 0.60 BCZT–0.40 CFO | 0-3 | 1.1 | 3.35 | [26] |
| 0.70 SBN–0.30 CFO | 0-3 | 3.5 | 0.037 | [35] |
| 0.55 BST-0.45 CFO | 0-3 | 1.8 | 0.4 | [36] |
| 0.70 PZT-0.30 NZFO | 0-3 | -- | 0.64 | [37] |
| BT- CFO | Core-shell | -- | 1.48 | [38] |
| CFO - BT | Core-shell | 1.4 | 2 | [39] |
| 0.70 BCTZ-0.30 CFO | 2-2 | 1.25 | 60 | [40] |
| NFO-BT | 2-2 | -- | 18 | [41] |
| CFO–PZT | 1-3 | -- | − 20 Oe m MV$^{-1}$ (Converse ME) | [42] |

*Conclusion*

To conclude, magnetoelectric (1-x) BCTSn – (x) CFO lead-free (0-3) particulate composites have been successfully synthesized by mechanical mixing of the individual ferroic phases. The composites' structure, ferroelectric, magnetic, and magnetoelectric properties were studied. XRD, Raman and SEM analysis confirm the coexistence of perovskite BCTSn and spinel CFO phases and phase purity. Ferroelectric and magnetic hysteresis loop measurements demonstrated that the BCTSn-CFO lead-free composites have multiferroic characteristics at room temperature. The largest ME coefficient is 0.1 mv cm$^{-1}$ Oe$^{-1}$ measured in 0.7BCTSn-0.3CFO composition, which is close to pseudo- cubic symmetry and relatively dense, reducing lattice strain and hence promoting domain switching. The obtained magnetoelectric results of the BCTSn-CFO composites might be improved by varying the sizes and grain morphology in the FE and FM phases, as well as the design of other nanostructured couplings between the magnetic and ferroelectric components.